\documentclass[conference,a4paper]{IEEEtran}

\usepackage{float}
\usepackage{cite}
\usepackage{combelow}
\usepackage{amsmath,amssymb,amsfonts}
\usepackage{xurl}
\usepackage{hyperref}
\usepackage{cleveref}
\usepackage{algorithmic}
\usepackage{graphicx}
\usepackage{textcomp}
\usepackage{xcolor}
\usepackage{booktabs}
\usepackage{threeparttable}
\usepackage{pifont}
\usepackage[numbers]{natbib}
\usepackage{textcomp}
\usepackage{enumitem}
\usepackage{pgfplots,pgfplotstable}
\pgfplotsset{compat=1.18}

\newcommand{\cmark}{\ding{51}}
\newcommand{\xmark}{\ding{55}}

\newcommand{\hvsockets}{Hyper-V sockets}

\ifdef{\anonymous}{
    \newcommand{\ph}[1]{\phantom{#1}}
}{
    \newcommand{\ph}[1]{#1}
}

\begin{document}

\title{Using Hyper-V Sockets for Real-time Data Extraction from a Malware Analysis Sandbox}

\author{\IEEEauthorblockN{\ph{Istv{\'a}n-Attila CS{\'A}SZ{\'A}R}}
\IEEEauthorblockA{\ph{\textit{Technical University of}} \\ \ph{\textit{Cluj-Napoca},} \\
\ph{\textit{Bitdefender}} \\
\ph{Cluj-Napoca, Romania} \\
\ph{istvan.csaszar@cs.utcluj.ro}}
\and
\IEEEauthorblockN{\ph{Radu-Marian PORTASE}}
\IEEEauthorblockA{\ph{\textit{Technical University of}} \\ \ph{\textit{Cluj-Napoca},} \\
\ph{\textit{Bitdefender}} \\
\ph{Cluj-Napoca, Romania} \\
\ph{rportase@bitdefender.com}}
\and
\IEEEauthorblockN{\ph{Adrian COLE\cb{S}A}}
\IEEEauthorblockA{\ph{\textit{Technical University of}} \\ \ph{\textit{Cluj-Napoca},} \\
\ph{Cluj-Napoca, Romania} \\
\ph{adrian.colesa@cs.utcluj.ro}}
\and
\IEEEauthorblockN{\ph{Adrian GROZA}}
\IEEEauthorblockA{\ph{\textit{Technical University of}} \\ \ph{\textit{Cluj-Napoca},} \\
\ph{Cluj-Napoca, Romania} \\
\ph{adrian.groza@cs.utcluj.ro}}
}

\maketitle

\begin{abstract}
We present how Hyper-V sockets can be used as a real-time communication channel for a malware analysis sandbox.
We show that, compared to WinSock TCP sockets, Hyper-V sockets are not subject to TCP/IP-layer blocking and are 
not enumerated by common TCP connection listing tools. We compare the throughput of the two communication channels
as a function of buffer size.
\end{abstract}


\section{Introduction}

According to the AV-Test institute, more than 120.000 new Windows malware samples registered every day \cite{avatlas}.
Without automated analysis, researchers have no chance of keeping up with the constant stream of new malware.
Malware sandboxes have a critical role in analyzing today's huge amount of malware, allowing researchers to
quickly triage new samples and focus their attention on the ones which present novel behavior and/or lack detection.

A malware analysis sandbox provides an isolated environment in which malware samples can be executed to record their behavior for analysis.
Examples of recorded behavior include processes created, modules loaded,
file system activity (files opened, created, read, written, etc.), registry activity (Windows specific), network activity and
relevant system APIs called.

Ransomware continues to remain a significant threat, accounting for over 83\% of malware types in ENISA's 2025 report \cite{enisathreatlandscape}.
Due to the nature of ransomware, analyzing it in a sandbox presents some unique challenges:
encryption of trace files and system tampering.

Since the main goal of ransomware is to encrypt files, there is a significant chance that the behavior trace files that
are saved inside the virtual machine (i.e. to be extracted later) will be encrypted by the ransomware.
While trace files can be protected from encryption, for example, by blocking destructive file operations that target them,
the correct implementation of such protection mechanisms is not trivial.
If the sandbox runs the ransomware with an active internet connection, the ransomware might also exfiltrate the
contents of the trace files, which may be undesirable.

Recent trends also show that ransomware may tamper with the system in order to maximize the chances of successful encryption
and hinder recovery. Common techniques include: termination of backup and security services, using vulnerable drivers to terminate security solutions, forcing the system to reboot in safe mode, encrypting recovery or even boot partitions.
This means that a sandbox that relies on saving behavior traces files in the virtual machine and copying them post-execution
might experience data loss because of system tampering (i.e. the file transfer component of the sandbox is stopped by the ransomware).

We initially built a sandbox based on Hyper-V, a type 1 hypervisor that is part of the Microsoft Windows operating system.
To copy files to and from the sandbox virtual machine, we used Powershell Direct, a feature of Hyper-V which allows running
arbitrary Powershell commands on a Windows guest virtual machine (VM) without attaching a virtual network adapter \cite{powershell-direct}.

When running ransomware with the Powershell-based setup, we found that in some cases the ransomware encrypted the trace files, making them
unusable, and in others stopped the Powershell Direct service, making it impossible to retrieve the files.
Knowing that Powershell Direct works without relying on a network connection, we started to investigate the communication channel Powershell Direct uses and whether it could be used as a real-time communication method to retrieve events from the guest agent without writing them to disk.

The main insight is that Powershell Direct is implemented through a service running on each Windows guest VM that uses a
Hyper-V socket to communicate with a Powershell instance on the host.
Once we found out that the Hyper-V sockets are available for any application to use \cite{hyperv-integration-service}, we set out to investigate
if this could be a suitable method to implement communication between the guest agent and the host for a malware sandbox.
More precisely, we set out to evaluate the following aspects:

\begin{enumerate}[leftmargin=3\parindent]
    \item[$RQ_1$] Throughput - Can the connection sustain real-time communication between the guest and the host?
    \item[$RQ_2$] Visibility - What methods are available to detect Hyper-V socket connections?
    \item[$RQ_3$] Resilience - Can the connection be blocked, redirected or closed?
\end{enumerate}

The contributions are: 
\begin{itemize}
    \item We show how to build a malware sandbox using Hyper-V that does not rely on traditional data extraction methods
          (disk based or network based).
    \item We measure the throughput of Hyper-V sockets and WinSock TCP sockets as a function of the buffer size used to send data,
          showing that Hyper-V sockets are also suitable for real-time communication.
    \item We show how to detect the presence of Hyper-V and WinSock TCP based connections, and the effect of starting the agent as                        a PPL (Protected Process Light) on the visibility of sockets.
    \item We systematize known methods that can be used to disrupt WinSock TCP socket based communication and show that they don't
          affect Hyper-V sockets.
\end{itemize}

The remainder of the paper is structured as follows.
\Cref{sec:backround} presents the key aspects to consider when building a malware sandbox.
\Cref{sec:design} presents the design and key aspects of implementation.
\Cref{sec:evalutation} shows the results of the throughput, visibility, and resilience evaluations.
\Cref{sec:related-work} shows related work, comparing our solution with the most relevant papers.
\Cref{sec:conclusions} concludes the paper.
\section{Background}
\label{sec:backround}

In the vast majority of cases, the isolation required by a sandbox is achieved by using virtualization: a hypervisor is used to create
one or more virtual machines (VMs) in which the malware is placed and executed. For recording behavior, there are two
possible approaches: agent-based and agentless.

In the first case, an agent made up of one or more analysis/monitoring programs (sensors) is deployed in the VM
(i.e. besides the analyzed malware) to record the malware's behavior and send the recorded data to a data collector program,
which may either be inside the hypervisor or inside a trusted VM that only hosts data collector and analysis programs (i.e. is never used
to run malware).

The agentless approach is often based on Virtual Machine Introspection (VMI) \cite{Garfinkel2003AVM}, which means that the recording is performed
from the hypervisor, in a manner that is transparent to the malware running inside the analysis virtual machine.

While VMI solutions have no agent that the malware can detect in order to evade the analysis in the sandbox
(i.e. by not performing malicious actions), they have to deal with the ``semantic gap'' problem:
the hypervisor can only monitor the guest activity at a virtualized hardware level (e.g., CPU instructions, memory reads/writes),
which need to be interpreted to extract the high-level behavior enumerated above.

The major advantage of agent-based solutions is that most virtualization systems (hypervisors) can be used to host the
VMs for the sandbox, allowing them to be built using commodity hardware and software. 
Agents can also use various techniques for monitoring actions \cite{automatedanalysis2008, dynamicaanalysis2019}:
function hooking, dynamic binary instrumentation (DBI), ETW (Event Tracing for Windows) and kernel mode drivers.

An important aspect which affects agent-based systems is how the execution data is obtained and analyzed: post-execution or real-time.

In the post-execution case, the agent in the analysis VM will write the data collected by the sensors to files inside
the VM, then, when the malware finishes running or is terminated after some amount of time,
the files will be copied to an outside location (e.g. a data collector VM or a storage server).
Files can be copied through a network connection, by using hypervisor specific mechanisms (e.g. shared folders in VMWare Workstation) or
by extracting the files from the virtual disk of the analysis VM (e.g. by attaching the virtual disk to a trusted
data collector VM).

Real-time data extraction requires an active connection between the agent in the malware analysis VM and a data collector application
on a trusted VM. Usually this is implemented as a network connection with a firewall between the analysis VM and the trusted VM
to prevent the malware from discovering or interacting with components of the sandbox.
While a sandbox may also use a hypervisor based communication method (such as hypercalls from a guest agent), implementing
such a system requires hypervisor support and great care needs to be taken to avoid vulnerabilities that can lead to VM
escape scenarios (i.e. the malware executing code on the host system).
\section{System design and implementation}
\label{sec:design}

\begin{figure}
    \centering
    \scalebox{0.7}{
        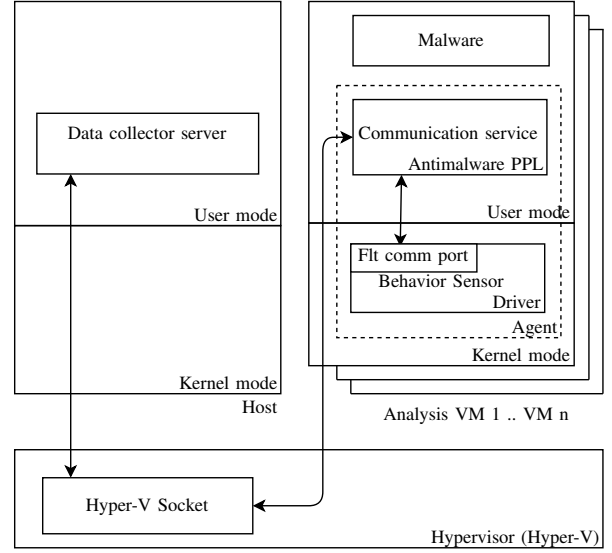
    }
    \caption{System architecture}
    \label{fig:architecture}
\end{figure}

The system consists of three components: (i) behavior sensor,
(ii) communication service, and (iii) data collector server (\autoref{fig:architecture}).

The \textit{behavior sensor} is implemented as a kernel mode filesystem minifilter driver that runs inside each VM to monitor actions of the analyzed malware. The driver can send messages to the communication service
through filesystem minifilter communication ports \cite{fltports}. To avoid the overhead of sending each event individually,
one message will contain multiple events. The number of events in each message is based on the total size (in bytes) of the resulting message.

The \textit{communication service} is implemented as user mode Antimalware PPL service that runs on each virtual machine and forwards messages
from the behavior sensor to the data collector server through a Hyper-V socket.
    
The \textit{data collector server} runs on the host and listens through a Hyper-V socket for connections
from the communication service in each VM.
As messages with behavior traces arrive, the server will persist them, e.g., as files or by sending them to a database.
The server can also send commands to the communication service inside the virtual machines (which may, in turn, be forwarded to the
behavior sensor), e.g. to request all events that were not yet sent.

Using an out-of-band communication method allows running malware in the sandbox with network access without the agent
and malware ``sharing'' the communication channel (i.e. the network adapter). This eliminates the need to filter agent-generated traffic
from the network traffic capture and eliminates the risk of agent-generated traffic being modified or blocked by the malware or by a firewall.
Thus a setup similar to the one described in \cite{2021curator} can be easily replicated,
simply redirecting all traffic through TOR, without worrying about the traffic from the guest agent.

The system also includes a monitoring mechanism that considers a virtual machine permanently compromised if the
connection is closed, causing the machine to be stopped by the hypervisor.
Subsequently the contents of the associated virtual disk can either be discarded (i.e. by reverting the virtual disk to a known clean state) 
or further analyzed for events that were generated after the connection was closed (i.e. by mounting the disk in separate virtual machine
that is in a known clean state and reading the files).

From the implementation view, Hyper-V sockets use a GUID based addressing scheme: an address consists of a \texttt{VmId} GUID and a \texttt{ServiceId} GUID.
\texttt{VmId} identifies a virtual machine or is a special wildcard and \texttt{ServiceId} identifies
the integration service using the Hyper-V socket. Each \texttt{ServiceId} GUID needs to be registered on the host system.
Drawing a comparison to TCP, \texttt{VmId} could be seen as the IP address and \texttt{ServiceId} could be seen as the port.

It should also be noted that Hyper-V sockets can only be used for communication between the host operating system (i.e.
the Windows operating system that was initially installed on the physical computer) and guest virtual machines.

We decided to implement the communication service in user mode to avoid adding more complexity to the kernel mode driver.
While normal sockets can be used from kernel mode, it requires more care and a more complex implementation than from user mode.
Since this service is in user mode, the malware could simply terminate the process,
or use other techniques such as code injection to tamper with the process.
To mitigate this problem, we configure each virtual machine to allow test signing and start the
communication service as an Antimalware Protected Process Light (PPL),
which is protected by the operating system, preventing other processes from terminating or performing
code injection via virtual memory writes or remote threads into the process \cite{ppl}.
It should be noted that while it is outside the scope of this paper,
malware could still tamper with the agent using a vulnerable driver to terminate the protected process or to remove kernel mode callbacks.

\section{Evaluation}
\label{sec:evalutation}
The proposed system is assessed against the three research questions: (i) throughput, (ii) visibility, (iii) resilience.
\subsection{Real time guest-host communication }

\pgfplotstableread[col sep=comma, header=true]{data/win11-hvsock.csv}\datavmserverhvsock
\pgfplotstableread[col sep=comma, header=true]{data/win11-native.csv}\datavmserverative
\pgfplotstableread[col sep=comma, header=true]{data/win10-hvsock.csv}\datadesktophvsock
\pgfplotstableread[col sep=comma, header=true]{data/win10-native.csv}\datadesktopnative

To assess the feasibility of \hvsockets, we measured their throughput compared to native sockets.
We consider throughput a decisive factor, since it defines if real-time communication can be achieved or not.

Another critical aspect related to throughput is whether it is influenced by the size of the messages, which is
determined by the number and size of the events included in each message. The number of events that are needed to constitute a message
also directly determines the latency of the communication.
Sending individual events is effectively real-time, while sending events in batches means that a certain number of events have to be collected before sending a message.

\begin{figure*}[htb]
\centering
\begin{tikzpicture}[baseline]
\begin{axis}[
    height=4.5cm,
    width=8cm,
    xlabel={Buffer Size (bytes)},
    ylabel={Throughput (MB/sec)},
    ymax=8000,
    xmode=log,
    log basis x={2},
    xtick=data,
    xticklabels={64, 128, 256, 512, 1k, 2k, 4k, 8k, 16k, 32k},
    grid=both,
    legend pos=north west,
    title={Throughput vs buffer size (System 1)},
]
    \addplot+ [
        error bars/.cd, y dir=both, y explicit
    ] table [
        x=BufSize,
        y=ThroughputMean,
        y error plus expr=\thisrow{ThroughputMax}-\thisrow{ThroughputMean},
        y error minus expr=\thisrow{ThroughputMean}-\thisrow{ThroughputMin}
    ]\datadesktophvsock;
    \addlegendentry{HVSock}
    \addplot+ [
        error bars/.cd, y dir=both, y explicit
    ] table [
        x=BufSize,
        y=ThroughputMean,
        y error plus expr=\thisrow{ThroughputMax}-\thisrow{ThroughputMean},
        y error minus expr=\thisrow{ThroughputMean}-\thisrow{ThroughputMin}
    ]\datadesktopnative;
    \addlegendentry{WinSock}
\end{axis}
\end{tikzpicture}%
~
\begin{tikzpicture}[baseline]
\begin{axis}[
    height=4.5cm,
    width=8cm,
    xlabel={Buffer Size (bytes)},
    ylabel={Throughput (MB/sec)},
    ymax=8000,
    xmode=log,
    log basis x={2},
    xtick=data,
    xticklabels={64, 128, 256, 512, 1k, 2k, 4k, 8k, 16k, 32k},
    grid=both,
    legend pos=north west,
    title={Throughput vs buffer size (System 2)}
]
    \addplot+ [
        error bars/.cd, y dir=both, y explicit
    ] table [
        x=BufSize,
        y=ThroughputMean,
        y error plus expr=\thisrow{ThroughputMax}-\thisrow{ThroughputMean},
        y error minus expr=\thisrow{ThroughputMean}-\thisrow{ThroughputMin}
    ]\datavmserverhvsock;
    \addlegendentry{HVSock}
    \addplot+ [
        error bars/.cd, y dir=both, y explicit
    ] table [
        x=BufSize,
        y=ThroughputMean,
        y error plus expr=\thisrow{ThroughputMax}-\thisrow{ThroughputMean},
        y error minus expr=\thisrow{ThroughputMean}-\thisrow{ThroughputMin}
    ]\datavmserverative;
    \addlegendentry{WinSock}
\end{axis}
\end{tikzpicture}
\caption{Throughput of WinSock TCP sockets and Hyper-V sockets on the two systems}
\label{fig:socket-perf}
\end{figure*}
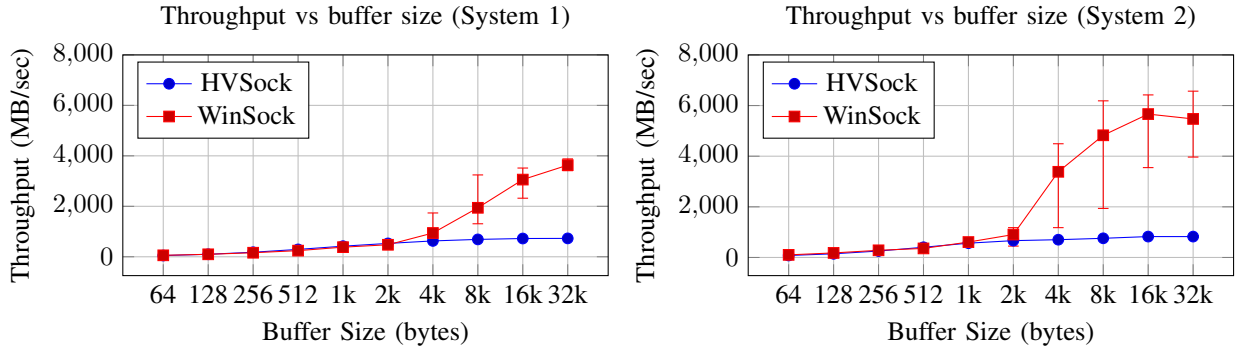

We performed benchmarks on two hosts, one with Windows 10 and one with Windows 11.
Their specifications are given in \cref{tab:specs}.
We deployed a virtual machine with Windows 10, 4 processors and 8GB of memory on each host. 
On each host, we tested the throughput for various message (buffer) sizes between 64 bytes and 32 kilobytes used to send and receive data between the host and the virtual machine.
We took 20 consecutive measurements, measuring the number of bytes transferred over 30 seconds from a sender (client) running inside the virtual machine and a receiver (server) running on the host.
The mean of the 20 measurements for each buffer size is shown \Cref{fig:socket-perf}, with error bars showing the minimum and
maximum of each of the 20 measurements.

For native sockets, we used an Internal network, configured with the default settings, with manually assigned IP addresses and the connection profile set to ``Private'' on the host.

During the tests, we noticed high CPU usage on both systems, so we also added host specific wait period (at most 2 minutes) to ensure the CPU is not thermally throttled as the tests progress.

\begin{table}
    \centering
    \caption{System specifications}
    \begin{tabular}{lllrl}
        \toprule
        Name     & OS          & Processor & RAM        & Disk     \\ \midrule
        System 1 & Windows 10  & i7-10700K & 128GB DDR4 & SATA SSD \\
        System 2 & Windows 11  & i7-13700  & 64GB  DDR4 & NVME SSD \\
        \bottomrule
    \end{tabular}
    \label{tab:specs}
\end{table}

Based on the results in \Cref{fig:socket-perf}, up to a buffer size of 1024 bytes, the performance of Hyper-V and Windows sockets are
almost identical, but above 4096 bytes, Windows sockets are significantly faster: Hyper-V sockets reach a maximum of 700MB/sec,
while Windows sockets can reach up to 6GB/sec. We also noted the difference in peak throughput for WinSock TCP sockets between the
two systems (4GB/sec for System 1 and 6GB/sec for System 2), but did not investigate further.

As noted above, the message (buffer) size is also determines number and size of events that can be included in a single message sent from the
guest agent to the host. Sending each event as an individual message ensures low latency, while batching more events
into a message reduces the overhead from each send operation, but increases latency and the quantity of data lost if the
agent in the sandbox terminates unexpectedly.
Based on the results, sending small messages reduces the throughput significantly.
The ideal buffer size seems to be above 4096 bytes and can be chosen based on the latency requirements.

We also note that for buffer sizes above 4096 bytes the performance of Hyper-V sockets is much more stable than that of Windows
sockets, an effect which is more pronounced on System 2.

\subsection{Detecting Hyper-V socket connections}
Many modern malware are designed to detect and evade (by not running or by performing benign actions) sandbox environments \cite{lindorfer2011detecting}.
Thus, in an ideal scenario, sandbox components and mechanisms should not be easily detectable by malware.
We used 3 tools of varying specialization that can detect network connections to investigate the visibility of the communication channel.

\textit{netstat} is a tool that is shipped with Windows for listing network connections. It relies on the undocumented \texttt{NsiAllocateAndGetTable} function (which is used by the \texttt{GetExtendedTcpTable} function internally)
to obtain all TCP connections on the system.

\textit{System Informer} \cite{systeminformer} is an open source advanced system monitoring tool, with capabilities that include listing
running processes, active network connections, open handles, loaded modules and monitoring system resource usage.
To obtain TCP and UDP connections, it uses the \texttt{GetExtendedTcpTable}, \texttt{GetExtendedUdpTable},
\texttt{InternalGetBoundTcpEndpointTable} and \texttt{InternalGetBoundTcp6EndpointTable} functions from the \texttt{iphlpapi.dll} module.
To obtain active Hyper-V socket connections on the host, it uses the undocumented device \verb|\Device\HvSocketSystem\HvSocketControl|,
and issues undocumented control codes using \texttt{NtDeviceIoControlFile} to obtain active listeners
\verb|(0x21C01C)| and connections \verb|(0x21C020)|.

\textit{AfdSocketViewer} \cite{afdsocketviewer} is an open source tool for forensic investigation and debugging of Windows sockets.
The authors reverse engineered the AFD.sys (Ancillary Function Driver) driver to understand its inner workings and document its
interface. The resulting tool is available on github \footnote{\url{https://github.com/huntandhackett/AfdSocketViewer}}.
To list information about sockets, the tool duplicates each handle that belongs to a process of interest and checks
if the volume of the handle is \verb|\Device\Afd|, then it uses the undocumented IOCTLs to obtain information about each handle.

For each communication method, we started a server process on the host system and a client process inside a virtual machine
that connected to the server on the host. We then inspected the output of each of the 3 tools to see if can identify the
connection. We then repeated the tests, this time running the client process as an Antimalware PPL (Protected Process Light),
which, among other security features, prevents other processes from duplicating handles from the process,
limiting the information that can be obtained.

\begin{table}
    \centering
    \begin{threeparttable}
    \caption{Visibility of socket types}
    \begin{tabular}{cccc}
        \toprule
        Method      & netstat -ano    & System Informer  & AFDSocketViewer \\ 
                    & Console output   & Network tab      & Console output  \\ \midrule
        WinSock     & \cmark \tnote{1} & \cmark           & \cmark          \\
        WinSock PPL & \cmark \tnote{1} & \cmark           & \xmark          \\
        HVSock      & \xmark           & \xmark \tnote{2} & \cmark          \\
        HVSock PPL  & \xmark           & \xmark \tnote{2} & \xmark          \\
        \bottomrule
    \end{tabular}
    \begin{tablenotes}
        \item[1] Connection is listed, but only PID of owning process is shown
        \item[2] The connection can be seen from the host, using host-only Hyper-V APIs
    \end{tablenotes}
    \label{tab:socket-visibility}
    \end{threeparttable}
\end{table}

\Cref{tab:socket-visibility} shows a summary of the results. In the case of WinSock, the connection is visible using all 3 tools.
When we start the process as a PPL, we can start to see the limits of tools that rely on handles:
netstat can only show the PID of the owning process and AFDSocketViewer fails to obtain information about the socket because
it can't obtain (duplicate) handles from a PPL.
In the case of Hyper-V sockets, netstat can't obtain any information since it's limited to TCP and UDP connections, 
neither can System Informer, since it only implements querying Hyper-V socket connections from the host system.
AFDSocketViewer can obtain full information about the Hyper-V socket, provided that it can duplicate handles.

\subsection{Blocking, redirecting or closing the connection}

To evaluate the resilience of Hyper-V sockets, we started by collecting known attacks against WinSock TCP socket based
connections on Windows. It should be noted that most of the techniques were developed as red-teaming tools in order
to evade or tamper with security or EDR (Endpoint Detection and Response) software running on compromised endpoints.

\textit{Windows Firewall based blocking} involves adding firewall rules to block the connection.
The most prominent tool using this technique is EDRSandblast \cite{edrsandblast}.
    
\textit{WFP (Windows Filtering Platform) based blocking} uses the Windows Filtering Platform API to create filters
that block the connection. Compared to Windows Firewall, WFP is meant to be used programmatically to implement custom
traffic inspection or blocking applications.
The most prominent tool using this technique is EDRSilencer \cite{edrsilencer}.
    
\textit{Forced connection termination} makes use of the \verb|SetTcpEntry| function to forcefully change the TCP
connection state to \verb|MIB_TCP_STATE_DELETE_TCB|. In practice most tools implement the function inline, by
directly opening the NSI (Network Store Interface) device and using undocumented control codes to achieve the same effect.
This technique is implemented both by the TcpNsiKill \cite{tcpnsikill} and System Informer \cite{systeminformer},
both using the undocumented control code \verb|IOCTL_NSI_SET_ALL_PARAMETERS (0x120013)|.

\textit{Null routing} adds a custom routing entries that redirect the EDR traffic, usually to the
\verb|0.0.0.0| IP address, resulting in the traffic being dropped. The method is described in \cite{nullrouting}.

\textit{Name resolution policy table} can be used to resolve EDR communication domains to a local or invalid IP address,
preventing communication. This method is described in \cite{edrcommblock1}.

\textit{Hosts file blocking} is a well known method of overriding DNS resolution, achieving a similar result
as the previous method. This method is described in \cite{edrcommblock2}.

\begin{table}
    \centering
    \begin{threeparttable}
    \caption{Connection blocking methods}
    \begin{tabular}{lll}
        \toprule
        Ref.                              & Method                        & Layers                  \\ \midrule
        \cite{edrsandblast}               & Windows Firewall              & Binary Path \tnote{1,2} \\
        \cite{edrsilencer}                & Windows Filtering Platform    & Binary Path \tnote{1,2} \\
        \cite{tcpnsikill, systeminformer} & Forced connection termination & TCP connection          \\
        \cite{nullrouting}                & Null routing                  & IP                      \\
        \cite{edrcommblock1}              & Name resolution policy table  & DNS                     \\
        \cite{edrcommblock2}              & Hosts file                    & DNS                     \\
        \bottomrule
    \end{tabular}
    \begin{tablenotes}
        \item[1] Also supports blocking connections by IP, Port and Protocol
        \item[2] Only IP-based connections are blocked
    \end{tablenotes}
    \label{tab:connection-termination}
    \end{threeparttable}
\end{table}

\Cref{tab:connection-termination} shows a summary of the layers where each blocking method operates.

It should be noted that while both Windows Firewall and Windows Filtering Platform support blocking connections by IP, Port and Protocol,
their main advantage is being able to block all traffic from a process that has a certain binary path. This means
that it is sufficient to have a list of process names whose connections should be blocked, instead of obtaining and maintaining
a list of domain names or IP addresses of their communication endpoints.

Another key aspect is that none of the blocking methods interact directly with the processes (i.e. try to open the process or to duplicate handles) whose connections they are trying to block. This allows the methods to work even on protected processes.

Since Hyper-V sockets don't use the TCP/IP stack, they are implicitly not affected by the attacks enumerated above, even though they
use the same programming interface (API) as WinSock TCP sockets.

\section{Discussion and related work}
\label{sec:related-work}

In \cite{sandboxarchitectures2025} \citeauthor{sandboxarchitectures2025} present a classification of malware sandboxes and their architectures.

There exist many commercial and open source examples of both VMI based and agent based sandboxes:
Examples of VMI based systems include DRAKVUF \cite{lengyel2014drakvuf} (open source) and
VMRAY \cite{VMRAY} (commercial).
Examples of agent-based solutions include: Cuckoo \cite{Cuckoo} (open source),
ANY.RUN \cite{anyrun} (commercial),
Hybrid Analysis \cite{hybridanalysis} (commercial).

The most prominent open source sandbox is the Cuckoo sandbox. While the original repository was archived in 2021 \footnote{\url{https://github.com/cuckoosandbox/cuckoo}}, a fork, named Cuckoo 3 is currently maintained by
the Estonian Information System Authority\footnote{\url{https://cuckoo-hatch.cert.ee/}}.
The original used function hooking to achieve monitoring by injecting a dll into each monitored process,
Cuckoo 3 uses a closed source kernel mode driver.

Compared to our solution Cuckoo uses a network adapter and sockets to communicate with a result server (data collector)
on the host \cite{cuckoo-docs}.

The authors of \cite{2021curator} present the Curator sandbox, a system for building behavioral datasets for Windows malware.
The solution is based on VMWare Workstation, it monitors behavior using a kernel mode driver and it includes an
Active Directory (AD) server in its architecture, and runs samples with internet access redirected through TOR.
It extracts behavior traces in two stages: first the traces are copied from the sandbox virtual machine to a shared folder hosted by a storage server virtual machine, then, after enough traces are collected on the storage server, the VM is shut down,
its virtual disk is mounted in a control server VM and stored for further analysis.

Compared to our solution, Curator only extracts the results post-execution, risking data loss if the sandbox virtual machine
crashes or is shut down unexpectedly during execution. To prevent malware (i.e. ransomware) from tampering with the trace files,
the authors implemented basic self protection mechanisms, but no further details are given in the paper.

\section{Conclusions}
\label{sec:conclusions}

This paper studies the feasibility of using Hyper-V sockets as a data extraction method from a malware analysis sandbox.
We compare Hyper-V sockets to WinSock TCP sockets from three points of view: throughput, visibility and resilience.
While their peak throughput (700 MB/sec) is lower than that of TCP sockets (6GB/sec), it is more stable and still sufficient
for real-time data transfer. Their visibility is also limited compared to TCP sockets, making them harder to detect.
We also studied attacks that are commonly used against WinSock based connections and concluded that they are not applicable to
Hyper-V sockets due to its GUID based addressing scheme.

\bibliographystyle{IEEEtranN}
\bibliography{aqtr2026}

\end{document}